# Gas relative permeability in unconventional reservoir rocks

Behzad Ghanbarian[1*], Feng Liang[2], and Hui-Hai Liu[2]

[1] Porous Media Research Lab, Department of Geology, Kansas State University, Manhattan 66506 KS, USA

[2] Aramco Services Company: Aramco Research Center—Houston, 16300 Park Row, Houston 77084 TX, USA

[*] Corresponding author's email address: ghanbarian@ksu.edu

**Abstract**

Accurate modeling of gas relative permeability ($k_{rg}$) has practical applications in oil and gas exploration, production and recovery of unconventional reservoirs. In this study, we apply concepts from the effective-medium approximation (EMA) and universal power-law scaling from percolation theory. Although the EMA has been successfully used to estimate relative permeability in conventional porous media, to the best of our knowledge, its applications to unconventional reservoir rocks have not been addressed yet. The main objective of this study, therefore, is to evaluate the efficiency of EMA, in combination with universal power-law scaling from percolation theory, in estimating $k_{rg}$ from pore size distribution and pore connectivity. We presume that gas flow is mainly controlled by two main mechanisms contributing in parallel: (1) hydraulic flow and (2) molecular flow. We then apply the EMA to determine effective conductances and, consequently, $k_{rg}$ at higher gas saturations ($S_g$), and the universal scaling from percolation theory at lower $S_g$ values. Comparisons with two pore-network simulations and six experimental measurements from the literature show that, in the absence of microfractures, the proposed model estimates $k_{rg}$ reasonably well in shales and tight porous rocks. More specifically, we found that the crossover point – gas saturation ($S_{gx}$) at which transport crosses from percolation theory to the EMA – is non-universal. The value of $S_{gx}$ is a function of pore space characteristics such as pore size distribution broadness and critical gas saturation. This means that one should expect $S_{gx}$ to vary from one rock sample to another.

**1. Introduction**

Unconventional reservoirs have been successfully explored and produced not only in the United States and North America but also in China [1,2]. Accordingly, they became one of the major contributors to energy supplies. Although research on fluid flow in tight and ultra-tight porous rocks has made significant progress in the past decade, there is still a long way to fully understand mechanisms and key factors/parameters that control gas transport in such formations.

Gas relative permeability ($k_{rg}$) is one of the crucial parameters to estimate gas production [3,4] or to evaluate the performance of $CO_2$ or miscible gas enhanced oil recovery (EOR) [5–7]. Although literature on single-phase gas permeability in unconventional reservoir rocks is extensive and vast [8–15], studies on gas relative permeability are still very limited. In what follows, we summarize some applications from pore-network simulations and theoretic approaches to the modeling of two-phase flow in unconventional reservoir rocks.

*1.1. Pore-network simulations*

Huang et al. [16] are among the first who investigated two-phase flow in shales. In their pore-network model, two wettability conditions – completely water-wet and mixed-wet – were considered. In the case of mixed-wet condition, the hydrocarbon-wet organic matter was embedded within the water-wet inorganic shale matrix. Huang et al. [16] incorporated the effects of slip flow and compressibility throughout their pore-network modeling, while the influence of gas sorption was only considered for hydrocarbon-wet (organic) pores. Their study started with the simulation of two-phase flow under completely water-wet conditions. In this case, some physical mechanisms, such as gas sorption, slip flow and gas compressibility were not considered. They found that although the relative permeability curves for the boundary pressure differences 0.05 and 5 MPa (7.25 and 725 psi) were similar, the curves for 5 and 500 MPa (725 and 72,500 psi) cases were very different, especially at low and intermediate gas saturations (see their Fig. 12). Huang et al. [16] stated that, "Calculations of the capillary numbers in the three cases indicate that the 0.05 and 5 MPa cases are dominated by capillary forces, while the 500 MPa case is dominated by viscous forces." As Huang et al. [16] pointed out, modeling two-phase flow under mixed-wet conditions is more complicated than that under completely water-wet circumstances. What makes the former more challenging is that while water imbibes into water-wet pores, its invasion in the hydrocarbon-wet organic region is a drainage process. In the absence of gas sorption, slip flow and compressibility, Huang et al. [16] compared the relative permeability curves of the mixed-wet and completely water-wet pore networks and found that $k_{rg}$ was consistently higher in the mixed-wet network (see their Fig. 13). This may be due to the presence of hydrocarbon-wet pores. They also investigated two-phase flow in the mixed-wet pore network while considering the physical mechanisms such as gas sorption, compressibility and slip flow. In their Fig. 14, Huang et al. [16] compared the gas relative permeability curves of the mixed-wet network with and without consideration of physics such as gas sorption, slip flow and compressibility. As presented in their Fig. 14, the gas relative permeability curves are very similar for these two cases, especially at intermediate to high gas saturations. However, discrepancies become more evident at low gas saturations. In the presence of gas sorption, slip flow and compressibility, the critical gas saturation is much lower. Huang et al. [16] concluded that, "Although the Klinkenberg [slip flow] effect is included in the model, its effect cannot be seen in the relative permeability curves. … Overall, the results shown in Fig. 14 indicate that additional invasion of the fracking fluid into the shale matrix can be expected in the field due to the various physics of the gas phase in the system."

Song et al. [17] investigated two-phase permeability of three-dimensional pore-network models constructed based on two-dimensional (2D) scanning electron microscopy (SEM) images of an organic-rich shale sample. They found that, "The results indicate that the influences of pore pressure and temperature on water phase relative permeabilities are negligible while gas phase relative permeabilities are relatively larger in higher temperatures and lower pore pressures. Gas relative permeability increases while water relative permeability decreases with the shrinkage of pore size. This can be attributed to the fact that gas adsorption layer decreases the effective flow area of the water phase and surface diffusion capacity for adsorbed gas is enhanced in small pore size."

In another study, Wang and Sheng [18] studied relative permeability using 2D pore-network models, for a Barnett shale sample, at pore pressures $p$ = 10 and 20 MPa (1,450 and 2,900 psi). Those authors defined two types of gas relative permeability $k_{rg}$: (1) Darcy permeability based $k_{rg}$ (their Eq. 11), and (2) normalized $k_{rg}$ where non-Darcy permeability was used to normalize permeability (their Eq. 12). Wang and Sheng [18] found that the normalized relative permeability did not vary much when the second definition of non-Darcy gas flow was used to normalize gas permeability. However, their results were based on two-dimensional simulations.

*1.2. Theoretical approaches*

In addition to the pore-network simulations, models based upon bundle of capillary tubes have also been developed to study two-phase permeability in tight and ultra-tight porous rocks. For instance, Zhang et al. [19] presumed that nano-scale inorganic pores in shales follow fractal scaling and their size distribution conforms to a power-law probability density function. They proposed a gas relative permeability model by considering effects of gas slippage in the entire Knudsen regime, stress dependence, and water film thickness. Zhang et al. [19] found that Knudsen diffusion and slippage effects boosted the gas relative permeability considerably at low to intermediate pressures (e.g., 0.1-1 MPa). They also stated that as pore space or tortuosity fractal dimension increased, gas relative permeability slightly increased, while water relative permeability decreased (see their Fig. 10).

Liu et al. [20] proposed power-law functions to model gas relative permeability in unconventional reservoir rocks based on the bundle of capillary tubes approach. Although theoretically developed, their models (see their Eqs. (31) and (32)) assume that gas relative permeability vanishes at zero gas saturation meaning that critical gas saturation is zero. By fitting their one-parameter model i.e., $k_{rg} = (1 - S_w)^\gamma$ to four experimental data, Liu et al. [20] found $\gamma$ ranged between 3.9 and 40.9.

*1.3. Objectives*

Although theoretical models were proposed for the $k_{rg}$ estimation, most of them are based on a "bundle of capillary tubes" concept, a severely distorted idealization of ultra-tight porous rocks that ignores the effect of pore connectivity. Alternatively, one may apply the effective-medium approximation (EMA), which has been successfully applied to estimate pore-pressure-dependent gas permeability in shales [13] and water relative permeability in soils [21]. However, it has never been used to model gas relative permeability in unconventional reservoir rocks. Accordingly, the main objectives of this study are to: (1) invoke concepts from the effective-medium approximation, (2) consider hydraulic flow and molecular flow as two main mechanisms contributing to gas flow in parallel, (3) scale up gas relative permeability $k_{rg}$ involving pore size distribution *f(r)* and pore connectivity derived from experimental measurements, and (4) compare the estimated $k_{rg}$ with both experimental measurements and simulation results. In the following sections, we first introduce the concept of effective-medium approximation. Then, we discuss gas transport in unconventional reservoir rocks and its mechanisms at the pore-scale level.

Next, we develop a model to estimate gas relative permeability from pore scale property comprising parameters of pore size distribution and pore connectivity.

**2. Effective-medium approximation**

The effective-medium approximation (EMA) is an upscaling technique from statistical physics, which can be used to model flow and transport in relatively heterogeneous systems. This method, originally developed by Bruggeman [22] and independently by Landauer [23], was successfully applied in the literature to study various petrophysical properties [13,24–30].

Within the EMA framework, a heterogeneous medium with a specific pore conductance distribution ($f(g)$) is replaced with a hypothetically homogeneous network (see Fig. 1). In the latter, all pores have an effective pore conductance $g_e$ whose value is determined by solving Eq. (1) [31]:

$$\int \frac{g_e - g}{g + \left[\frac{Z}{2} - 1\right] g_e} f(g) dg = 0 \qquad (1)$$

where $Z$ is the average pore coordination number. In Eq. (1), $2/Z$ represents the percolation threshold ($p_c$) [32]. This means that the percolation threshold is a function of medium's pore connectivity. We should point out that although within bond percolation theory, $p_c = 1.5/Z$ in three dimensions, the effective-medium approximation predicts $p_c = 2/Z$ for both 2 and 3D networks.

In practice, $Z$ may be determined from either 3D images [33] or nitrogen sorption isotherms [34]. However, one may approximately set $Z = 2/S_{gc}$ in which $S_{gc}$ represents the critical gas saturation for percolation [26,32]. Accordingly, Eq. (1) changes to

$$\int \frac{g_e - g}{g + \left[\frac{1 - S_{gc}}{S_{gc}}\right] g_e} f(g) dg = 0 \qquad (2)$$

This approximation, Eq. (2), has been successfully evaluated for the estimation of water relative permeability in soils [21] as well as single-phase gas permeability in shales [13] and tight-gas sandstones [35].

**3. Gas transport in unconventional reservoir rocks**

One of the key parameters for gas flow in nano-scale pores is the Knudsen number ($Kn$), the ratio of the mean free path to the pore characteristic length. Different flow regimes can be defined based on the value of Knudsen number such as continuum ($Kn \leq 10^{-2}$), slip flow ($10^{-2} < Kn < 0.1$), transition flow ($0.1 < Kn < 10$) and molecular flow ($Kn \geq 10$) [36]. Accordingly, modeling relative permeability in tight and ultra-tight rocks is challenging because gas transport is affected by the complex physical mechanisms such as slip flow, Knudsen diffusion, sorption, and surface diffusion. Moreover, the pore space, where the gas predominantly remains in mudrocks, consists of nano- to micro-scale pores in organic patches and inorganic matrix with various wettability characteristics. In addition to that, the presence of microfractures in the organic patches [37] or inorganic matrix [38] will introduce higher level of complexity to gas transport modeling in these types of rocks. In the following, we assume that hydraulic flow and molecular flow are two main mechanisms contributing in parallel to gas transport. We first describe each mechanism at the pore scale. Next, we apply the EMA to develop a model and estimate gas relative permeability for a network of pores comprising parameters of pore size distribution and pore connectivity. The latter is reflected in the critical gas saturation $S_{gc}$, as described above.

*3.1. Hydraulic flow in a cylindrical pore*

Non-slip boundary condition may not accurately capture the physics of flow, particularly at the nano- and micro-scale levels [36]. In unconventional reservoir rocks, slippage of gas molecules on surface walls facilitates gas flow, and its effect should be incorporated into the modeling of gas transport. The hydraulic conductance ($g_h$) of a cylindrical pore of radius $R$ and length $l$ filled with a gas of viscosity $\mu$ is described in Eq. (3) [9,39]:

$$g_h = F \frac{\pi R^4}{8\mu l} \tag{3}$$

in which $F$ is the dimensionless coefficient for slippage and is given by [39]

$$F = 1 + \left(\frac{8\pi R_g T}{M_m}\right)^{0.5} \frac{\mu}{pR}\left(\frac{2}{TMAC} - 1\right) \tag{4}$$

where $M_m$ is the molar mass, $R_g$ is the universal gas constant, $T$ is the temperature, $p$ is the pore pressure, and TMAC is the tangential momentum accommodation coefficient. Agrawal and Prabhu [40] proposed an empirical relationship to estimate TMAC from Knudsen number $K_n$. However, it returns negative TMAC values for $K_n > 23.1$. For the sake of simplicity, we approximately set TMAC = 0.8 [9,36].

Ghanbarian and Javadpour [13] employed Eq. (3) in combination with Knudsen diffusion and the EMA to model single-phase gas permeability in shales and found good match between theory and measurement/simulation.

*3.2. Molecular flow in a cylindrical pore*

When the mean free path of gas molecules is of the same order or greater than the pore diameter, the majority of molecules move along straight trajectories until hitting a wall. Under the range of relatively large Knudsen number conditions, gas molecules move independently of each other and collisions between the molecules happen rarely. This type of flow called molecular is only caused by the kinetic energy of the molecules [41,42]. The molecular conductance ($g_m$) in a cylindrical pore of radius $R$ and length $l$ is [41,43]

$$g_m = \alpha \pi R^2 \sqrt{\frac{R_g T}{2\pi M_m}} \qquad (5)$$

in which $\alpha$ is

$$\alpha = 1 + y^2 - y\sqrt{1+y^2} - \frac{\left[(2-y^2)\sqrt{1+y^2}+y^3-2\right]^2}{4.5y\sqrt{1+y^2}-4.5\ln(y+\sqrt{1+y^2})} \qquad (6)$$

where $y = l/2R$.

*3.3. Total gas flow in a cylindrical pore*

Assuming that gas flow is mainly controlled by hydraulic flow and molecular flow (two mechanisms contributing in parallel) in a single cylindrical tube with radius $R$ and length $l$, the total conductance ($g_t$), at the pore scale is given by

$$g_t = g_h + g_m = F\frac{\pi R^4}{8\mu l} + \alpha \pi R^2 \sqrt{\frac{R_g T}{2\pi M_m}} \qquad (7)$$

Although the linear superposition of transport mechanisms i.e., molecular flow and hydraulic flow employed here to determine the total gas conductance is a first-order approximation, it has been previously validated in the literature to be accurate enough (see [44] and references therein).

*3.4. Gas relative permeability*

In this section, we scale up gas transport in a network of pores from pore-scale properties e.g., pore size distribution and pore connectivity. Although power-law and log-normal probability density functions were frequently used to fit the size distribution of pores in rocks, in this study, we use the actual pore size distribution and derive the total pore conductance distribution from $f(g_t)dg_t = f(r)dr$.

We further presume that each pore is occupied by either water (or oil) or gas, and all pores with size *r* and greater are accessible to gas. If capillary pressure curve is not available, the gas saturation, $S_g$, corresponding to pore radius *r* can be determined by integrating $r^3 f(r)$ between *r* and $r_{max}$ and normalizing that as follows

$$\frac{S_g - S_{gr}}{1 - S_{gr}} = \frac{\int_r^{r_{max}} r^3 f(r) dr}{\int_{r_{min}}^{r_{max}} r^3 f(r) dr} \tag{8}$$

where $S_{gr}$ is the residual gas saturation, formed from gas trapped in some subset of the pores in the range [$r_{min}$, $r_{max}$]. Note that since $S_g$ represents the volume fraction of gas, in Eq. (8) one needs to integrate $r^3 f(r)$ in which $r^3$ denotes the volume of a cylindrical pore whose length is linearly proportional to its radius *r* ($l \propto r$).

Under fully saturated conditions ($S_g = 1$), the effective conductance $g_e$ is given by

$$\int_{g_{tmin}}^{g_{tmax}} \frac{g_e(S_g=1) - g_t}{g_t + \left[\frac{1-S_{gc}}{S_{gc}}\right] g_e(S_g=1)} f(g_t) dg_t = 0 \tag{9}$$

where $g_{tmin}$ and $g_{tmax}$ are respectively the minimum and maximum total conductances in the pore network. Recall that $f(g_t)dg_t = f(r)dr$.

Under partially-saturated conditions, one has

$$\int_{g_{tmin}}^{g_t} \frac{g_e(S_g) - 0}{0 + \left[\frac{1-S_{gc}}{S_{gc}}\right] g_e(S_g)} f(g_t) dg_t + \int_{g_t}^{g_{tmax}} \frac{g_e(S_g) - g_t}{g_t + \left[\frac{1-S_{gc}}{S_{gc}}\right] g_e(S_g)} f(g_t) dg_t = 0 \tag{10}$$

$g_e(S_g = 1)$ and $g_e(S_g)$ are computed by numerically solving Eqs. (9) and (10), respectively. To determine the saturation dependence of gas permeability, $g_e(S_g)$ is calculated at various $S_g$ values.

Since gas permeability is proportional to the effective conductance (see Eqs. 35 and 36 in [25]), we define $k_{rg}$ as

$$k_{rg} = \frac{k_g(S_g)}{k_g(S_g=1)} = \frac{g_e(S_g)}{g_e(S_g=1)}. \tag{11}$$

Theoretically, one expects $k_{rg}$ to vanish as $S_g$ tends to its critical value, $S_{gc}$. It is well-known that the EMA is not accurate near the percolation threshold (or the critical gas saturation). Hence, we apply the EMA at high to intermediate gas saturations and use the universal power-law scaling from percolation theory at low gas saturations close to $S_{gc}$. The crossover between the two occurs at some gas saturation $S_{gx}$.

The universal power-law scaling from percolation theory is

$$k_{rg} = k_0(S_g - S_{gc})^t, \qquad S_g > S_{gc} \tag{12}$$

where the scaling exponent $t = 1.3$ and 2 in two and three dimensions, respectively, and $k_0$ is a numerical prefactor. Eq. (12) is called universal power-law scaling because the value of the exponent $t$ is universal and does not vary from one medium to another. Its value only depends on the dimensionality of the system [32].

The value of $k_0$ is determined by setting Eqs. (11) and (12) equal at $S_{gx}$ for which one needs to set the first derivative of the two equations equal. Since Eq. (11) is an implicit function of $S_g$, we numerically compute the values of $k_0$ and $S_{gx}$ as follows. We first interpolate the estimated $k_{rg}$ values at numerous gas saturations between $S_{gc}$ and 1 using the spline method. We then calculate $k_0$ and the slope at each point. $S_{gx}$ corresponds to the gas saturation at which the slopes of Eqs. (11) and (12) are equal. However, for a sample with a large $S_{gc}$ value, one may not find a smooth crossover point. In such a case, the value of $S_{gx}$ may be determined from a gas saturation at which the difference in slopes is minimal.

Accordingly, one should expect a sharp transition from the percolation theory scaling to the EMA one.

It is noted that Eq. (12) has been successfully applied to model gas relative permeability over the entire range of gas saturation in conventional porous media [32,45–47]. Eq. (12) was also combined with the linear scaling law from the EMA by Ghanbarian et al. [48,49] to model diffusion in porous materials. Daigle et al. [50] and Ojha et al. [51] used such a model to describe relative permeability in shale samples. However, neither Daigle et al. [50] nor Ojha et al. [51] compared model estimations with experimental measurements or simulation results. In unconventional reservoir rocks where gas permeability is pore pressure-dependent, one should use the universal power-law scaling with caution to model/estimate gas relative permeability over the entire range of gas saturation. This is because the experimental observations of Li et al. [52] and numerical simulations of Song et al. [17] indicate that as pore pressure decreases gas relative permeability should increase.

## 4. Materials and Methods

The database used in this study to evaluate our gas relative permeability model, Eq. (11) in combination with Eq. (12), consists of six experiments from references [53–57] and two pore-network simulations from references [17,58]. In the following, each dataset is briefly described, and detailed information could be found from the original publications.

*4.1. Experimental datasets*

*- Cases I & II: Bennion and Bachu (2007)*

Bennion and Bachu [53] collected two shale samples from central Alberta, Canada, namely Calmar shale and Colorado shale. The pore size distributions were derived from the measured mercury intrusion porosimetry curves (see their Fig. 1). For Case I, Calmar shale, they reported a narrow range of pores e.g., $1.5 \leq r \leq 9$ nm, while for Case II, Colorado shale, pore size spans more than one order of magnitude ($1.5$ nm $\leq r \leq 4.2$ $\mu m$) indicating a broader pore size distribution compared to Calmar shale. To measure $CO_2$ relative permeability, Bennion and Bachu [53] first evacuated samples to remove all trapped gas. They applied net reservoir pressure and then saturated samples with brine. Samples were heated to reservoir temperature (i.e., 43°C) while maintaining pore and net overburden pressures. Those authors used the unsteady-state method with supercritical-phase water-vapor saturated $CO_2$ (primary drainage test) in the water saturated cores. Irreducible water and maximum $CO_2$ saturations as well as the endpoint permeability and relative permeability to $CO_2$ were measured. Bennion and Bachu [53] used a computer history-matching method to generate the $CO_2$ and brine relative permeability curves for the primary drainage and imbibition displacement tests for each sample. However, only the primary drainage results were used in this study. To estimate $k_{rg}$, for the Calmar shale, we set $S_{gc} = 0$, which was determined from the actual $k_{rg}$ measurements; while for the Colorado shale, $S_{gc} = 0.30$, which was estimated from the inflection point of the measured mercury intrusion capillary pressure (MICP) curve. In the Results section, we address that $S_{gc} > 0$ (estimated from the measured MICP curve) would cause non-trivial underestimations by our theoretic model for the Calmar sample.

*- Case III: Dacy (2010)*

Dacy [54] reported air-brine capillary pressure curve measured via centrifuge as well as gas relative permeability on a low-permeability sand sample with porosity of 0.175. The Klinkenberg-corrected gas permeability and liquid permeability values were respectively 2 and 1.06 mD (see his Fig. 3). Gas relative permeability was measured using the unsteady-state method (his Fig. 4). To determine gas relative permeability in this dataset, gas permeability values were normalized to that measured at water saturation $S_w \sim$ 0.65 meaning $k_{rg} = 1$ at $S_g = 0.35$. $S_{gc} = 0.08$ was determined from the mercury saturation corresponding to the inflection of the measured capillary pressure curve shown in his Fig. 3.

*- Case IV: Yassin et al. (2016)*

The unconventional reservoir rock from Yassin et al. [55] was a Montney tight gas siltstone sample from the Western Canadian Sedimentary Basin. Those authors collected five samples from upper Montney and four samples from lower Montney. The average porosity and permeability values were respectively 6.7% and 2.93 $\mu$D for the upper Montney and 3.93% and 1.72 $\mu$D for the lower Montney samples (see their Table 1). Pore size distribution was determined from MICP measurements with maximum injection pressure 400 MPa only on one sample (see their Fig. 3). The inflection point on the MICP curve corresponds to the mercury saturation $S_{Hg} = 0.26$. This is the minimum saturation required to form the sample-spanning cluster and let the system percolate. We accordingly set $S_{gc} = 0.26$ in this dataset.

Gas relative permeability was measured on three Montney samples. For this purpose, Yassin et al. [55] saturated the samples using simulated formation water. Using a centrifuge apparatus and spinning under air at incrementally increasing speeds samples were gradually desaturated. After reaching equilibrium, samples were removed from the centrifuge and wrapped in cellophane for water redistribution. Water saturation was then measured using the mass balance method. Finally, effective permeability of gas at various water saturations was determined via a pulse decay permeameter at 6.9 MPa (1000 psi) pore pressure.

*- Cases V & VI: Nazari Moghaddam and Jamiolahmady (2019)*

Samples from Nazari Moghaddam and Jamiolahmady [56,57] were an Eagle Ford and a Pierre shale, prepared by cutting the samples in parallel to the bedding plane. MICP curves were measured using pressures ranged between 0 and 55,000 psi reported by Nazari Moghaddam and Jamiolahmady [59]. The pore size distributions were then determined from the measured MICP curves. To estimate $k_{rg}$ for the Eagle Ford, we set $S_{gc}$ = 0.13, which was estimated from the inflection point of the measured MICP curve [13]. However, mercury saturation corresponding to the inflection point for the Pierre sample was 0.35 [13]. This value resulted into significant $k_{rg}$ underestimation. To evaluate our theoretic model independent of how well $S_{gc}$ can be estimated, we set $S_{gc}$ = 0.08, which was determined from the actual $k_{rg}$ measurements for the Pierre sample.

To measure gas relative permeability, the samples were first dried overnight in the oven at 105 ºC while connected to a vacuum pump. Nazari Moghaddam and Jamiolahmady [56,57] measured the single-phase gas permeability by nitrogen ($N_2$) at full gas saturation. After that, the gas relative permeability was measured under liquid imbibition and steady-state conditions at pressure 10.3 MPa (1500 psi), as described in the following. The cores were first saturated by $N_2$, and then crude $C_4$ and $N_2$ were co-injected into the shale and continued to reach out the equilibrium state at several liquid-to-gas ratios. The equilibrium condition was achieved when such a ratio was equal to the rate of gas and liquid retractions. Pressures were measured at the inlet and outlet by a Quartz-dyne pressure transducer, and four Quizix pumps were simultaneously used for fluid injection and retraction.

*4.2. Pore-network simulations*

*- Case VII: Song et al. (2018)*

Song et al. [17] studied gas relative permeability of a three-dimensional pore network constructed from two-dimensional SEM images of an organic-rich shale sample. Those authors took into account the effect of various gas transport mechanisms such as viscous flow, Knudsen diffusion, adsorption as well as surface diffusion and applied invasion percolation to simulate gas relative permeability. The porosity of the pore network and the pore length to its radius ratio were 0.15 and 2.9, respectively (Wenhui Song, 2018; personal communication). Song et al. [17] simulated gas relative permeability at temperature $T = 127$°C and under a variety of pore pressures i.e., from 5 to 50 MPa (see their Fig. 8). In order to estimate gas relative permeability, we set $p = 25$ MPa, as an average value consistent with their simulations. Song et al. [17] also stated that 5% of the total pores were dead-end. As a rough approximation, we accordingly set $S_{gc} = 0.05$ for this dataset, since mercury intrusion porosimetry data are not available. As Song et al. [17] reported, their pore-network model included 400×400×400 voxels each of which was near 12 nm. Given that the geometric pore radius is 10.2 nm (determined from fitting a log-normal probability density function to the pore size distribution), their pore-network size should be large enough and above the representative elementary volume (REV).

*- Case VIII: Wang and Shang (2019)*

Wang and Sheng [58] simulated gas relative permeability in shale formations using a multi-scale model composed of cubic organic patches distributed within an inorganic pore network. Those authors generated a pore network with porosity of 0.075 (organic porosity = 0.035 and inorganic porosity = 0.04) and mean pore size of 110 nm. To mimic the actual pore size distribution in shales, Wang and Sheng [58] modified data from Berea Sandstone by shrinking its pore size 100 times (Xiukun Wang, 2019; personal communication). The MICP data are not available for this dataset. The simulated $k_{rg}$ curve, however, indicates that the critical gas saturation should be around 0.43. We accordingly set $S_{gc}$ = 0.43 to estimate $k_{rg}$ from the pore size distribution. The residual water saturation in their simulations was 0.2 ($S_{wr}$ = 0.2). We, therefore, normalized the estimated gas permeability using the value determined at $S_g$ = 0.8. Wang and Sheng [58] did not mention at which pore pressure their simulations were carried out. To estimate $k_{rg}$ for this dataset, the value of 6.9 MPa was assumed.

Salient properties of all datasets are summarized in Table 1. For all the experiments and simulations studied here, we set $S_{gr}$ = 0 to estimate $k_{rg}$. Following Patzek [60], we also set the ratio of pore-throat length to pore-throat radius equal to three ($l/R$ = 3), if not reported.

**5. Results**

In this section, we compare the estimated $S_g$-$k_{rg}$ curves with either experimentally measured or numerically simulated cases. For this purpose, we use natural scale plots instead of semi-log or log-log plots due to uncertainties in digitizing datapoints, particularly at low gas saturations near zero.

*5.1. Comparison with experimental measurements*

We first present the obtained results for the Bennion and Bachu [53] dataset. Fig. 2 displays the pore size distributions and $S_g$-$k_{rg}$ curves for both Calmar and Colorado shale samples. As shown, pore sizes span near one order of magnitude for Calmar shale, while more than three orders of magnitude for Colorado shale. This means that the latter is more complex in terms of pore-scale heterogeneity than the former. Accordingly, one should expect $k_{rg}$ for Colorado shale to be less than that for Calmar shale, demonstrated in Fig. 2.

The $k_{rg}$ estimations via the EMA (Eq. 11) and power-law scaling from percolation theory (Eq. 12) are also given in Fig. 2. Results of the EMA and universal power-law scaling from percolation theory are denoted respectively by blue and red lines. From the measured capillary pressure curve reported by Bennion and Bachu [53], see their Fig. 2, we found that the inflection point corresponds to some mercury saturation greater than zero. However, it is clear from the $k_{rg}$ measurements (Fig. 2) that $S_{gc} = 0$ for Calmar shale. Results (not shown) indicated that $S_{gc} > 0$ would cause $k_{rg}$ underestimation via our model. Although in the following we show that estimating $S_{gc}$ from the inflection point of mercury intrusion porosimetry curve results into accurate prediction of $k_{rg}$, further investigation is required to address alternative methods for estimating $S_{gc}$ in unconventional reservoir rocks.

Fig. 2 also shows results for the Colorado shale sample for which we found $S_{gc}$ = 0.3, determined from the inflection point of MICP curve. As can be seen, the estimated $k_{rg}$ curve agrees well with the measured data. For Colorado shale, we also found $S_{gx}$ = 0.87 – the crossover gas saturation at which the EMA scaling switches to the universal power-law scaling from percolation theory – nearly twice greater than that for Calmar shale ($S_{gx}$ = 0.43). The crossover point is non-universal and its value is a function of pore space characteristics such as pore size distribution broadness as well as percolation threshold (or equivalently here critical gas saturation) [32]. This means that one should expect $S_{gx}$ to vary from one rock sample to another.

For the Colorado shale sample with $S_{gc}$ = 0.3 (Fig. 2b), we found a sharp switch from the percolation theory scaling to the EMA one. As stated earlier, in this case $S_{gx}$ corresponds to the gas saturation at which the difference in slopes of Eqs. (11) and (12) is minimal. Ghanbarian et al. [49] also reported such an abrupt transition for gas relative permeability measurements in sand packs.

Fig. 3 shows the estimated $k_{rg}$ curves for the tight-gas sand sample from Dacy [54]. As can be observed, pore sizes span two orders of magnitude. Recall that in this dataset gas permeability values were normalized to that measured at $S_g$ = 0.35. Our results indicate that the proposed model, Eq. (11) in combination with Eq. (12), reduced to the universal power-law scaling from percolation theory shown in red in Fig. 3. We also found that estimating $S_{gc}$ from the mercury saturation corresponding to the inflection point on the MICP curve resulted into accurate estimation of $k_{rg}$ for this dataset.

Results for the tight-gas siltstone from Yassin et al. [55] are presented in Fig. 4. Pore radius ranges between 0.0023 and 0.24 µm, spanning near two orders of magnitude, and $r = 0.1$ µm corresponds to the mode of the pore size distribution. Comparison of the estimated $k_{rg}$ curve with the measured one obtained from three Montney samples are also shown in Fig. 4. We found that our model with $S_{gc} = 0.26$ estimated $k_{rg}$ accurately from the pore size distribution. Yassin et al. [55] developed a dual-wettability $k_{rg}$ model based on the bundle of capillary tubes approach and more specifically the Purcell [61] model. However, they overestimated $k_{rg}$ and had to replace the tortuosity factor $(1 - S_w)^2$ with $(1 - S_w)^3$ to obtain better match between the measured and estimated $k_{rg}$ (see their Figs. 9b-d). Interestingly, our model estimated $k_{rg}$ precisely without adjusting/tuning any parameters. We also emphasize that all parameters in our theoretic approach are physically meaningful. In Fig. 5, we present the obtained results for the Eagle Ford and Pierre samples from Nazari Moghaddam and Jamiolahmady [56,57]. Similar to the tight-gas siltstone sample shown in Fig. 4, the Eagle Ford and Pierre pore size distributions are negatively skewed indicating the non-trivial contribution of smaller pores in such unconventional reservoir rocks. Yassin et al. [55] also noticed a long tail at the lower end of the pore size distribution. For Eagle Ford shale, pore radius ranges between 0.002 and 0.2 µm, spanning two orders of magnitude. However, for Pierre shale, $r$ ranges between 0.002 and 0.08 µm, spanning less than two orders of magnitude.

Fig. 5 shows the estimated and measured gas relative permeabilities against gas saturation for the Eagle Ford and Pierre shale samples. Similar to the Colorado shale and tight-gas siltstone samples, the crossover gas saturation $S_{gx}$ is around 0.8 ($S_{gx} = 0.79$ and 0.84 respectively for Eagle Ford and Pierre).

As can be seen in Fig. 5, for Eagle Ford shale, $k_{rg}$ estimates are relatively accurate at low gas saturations. However, deviations between theory and experiment are observed at intermediate $S_g$ values. Overall, the proposed model, Eq. (11) in combination with Eq. (12), underestimated $k_{rg}$ for this case. This underestimation is in accord with the findings from Ghanbarian and Javadpour [13], who estimated single-phase permeability using the EMA on the same sample. Those authors found that the EMA underestimated gas permeability and attributed that to the presence of microfractures in the sample that were not captured via MICP. They stated that, "… the critical saturation in the Eagle Ford sample is remarkably less than that in the Pierre and Barnett samples, which might be an indication of microfractures in the Eagle Ford sample, associated with coring processes that are not reflected in the pore throat-size distribution." However, our theoretic model estimated $k_{rg}$ accurately over the entire range of gas saturation for Pierre shale. This means if $S_{gc}$ can be estimated with high accuracy and whether the pore size distribution can be determined precisely, our model would be able to estimate $k_{rg}$ accurately.

In most datasets studied here, $k_{rg}$ was experimentally measured at low to intermediate gas saturations e.g., $S_g < 0.5$. Therefore, model evaluation at high $S_g$ values are still required. We address the predictability of the proposed model, Eq. (11) in combination with Eq. (12), at high gas saturations using the pore-network simulations in the following section.

*5.2. Comparison with pore-network simulations*

In Fig. 6, we show the pore size distribution, estimated $k_{rg}$ curve at $p = 25$ MPa, and pore-network simulations from Song et al. [17]. This figure demonstrates that the proposed $k_{rg}$ model estimated gas relative permeability accurately over the entire range of gas saturation. More specifically, we found that $S_{gc}$ = 0.05, estimated from the fraction of dead-end pores, is in well agreement with the simulations.

Fig. 6 also shows that although the pore-network simulations vary by pore pressure, differences in $k_{rg}$ for various pore pressure values are not substantially remarkable, particularly at low and high gas saturations. This is consistent with two-dimensional simulations of Wang and Sheng [18] who reported similar $k_{rg}$ values for $p$ = 10 and 20 MPa (see their Figs. 7 and 8). Similar results were also theoretically obtained by Zhang et al. [19]. Huang et al. [16] also found that the relative permeability curves for the boundary pressure difference 0.05 and 5 MPa were similar. However, they reported that the curves for 5 and 500 MPa cases were different, particularly at low and intermediate gas saturations (see their Fig. 12). As Huang et al. [16] stated, this might be because at lower pore pressures gas transport is dominated by capillary forces, while at high $p$ values viscous forces dominate.

Fig. 7 shows the pore size distribution as well as the estimated and simulated gas relative permeability curves for the pore-network model of Wang and Sheng [58]. Pore size spans near one order of magnitude with a weighted arithmetic mean of 98 nm. This value is very close to that Wang and Sheng [58] reported (i.e., 110 nm). As can be seen, the EMA, shown in blue, estimated gas relative permeability at high gas saturations accurately. The universal power-law scaling from percolation theory, shown in red, also estimated $k_{rg}$ at low to intermediate $S_g$ values precisely. We should point out that the value of critical gas saturation for this dataset was not available. We, accordingly, had to determine its value ($S_{gc}$ = 0.43) from the simulated $k_{rg}$ curve.

## 6. Discussion

*6.1. Gas transport mechanisms*

Various models and approaches were developed to better describe gas flow in tight and ultra-tight porous rocks. Most take into account the effects of slip flow and Knudsen diffusion [8,9,13,62,63]. However, more recently the influences of surface diffusion and sorption were incorporated [64–68]. In this study, we only incorporated the effects of two mechanisms i.e., hydraulic flow and molecular flow to estimate gas relative permeability from pore space characteristics such as pore size distribution and pore connectivity. In practice, however, other mechanisms e.g., sorption and surface diffusion might non-trivially contribute to gas transport in shales and tight porous rocks, as addressed in several studies. For example, Song et al. [17] considered viscous flow, Knudsen diffusion, surface diffusion, sorption, and gas PVT and viscosity in their pore-network modeling. Reasonable comparison between our theoretic approach and their PNM simulations shows that the effect of sorption and surface diffusion is not substantial. In their pore network model, pore body and pore throat sizes ranged from 7 to 54 and 6 to 40 nm, respectively. Recently, Wu et al. [64] proposed a unified theoretical model by coupling slip flow, Knudsen diffusion, sorption, and surface diffusion. They stated that "… surface diffusion is an important transport mechanism, and its contribution cannot be negligible and even dominates in nanopores with less than 2 nm in diameter". Naraghi et al. [69], however, claimed that surface diffusion has trivial impact on total gas flow, based on 2D stochastic models of shale samples at 13.8 MPa (2000 psi) pore pressure. Remarkable discrepancies between 2D and 3D models, representing shales, and their gas permeability estimations have been recently highlighted by Cao et al. [70]. Those authors considered slip flow and Knudsen diffusion mechanisms and demonstrated that deviations between 2D and 3D models are negligible only if pore pressure is greater than 10 MPa. Interestingly, Finkbeiner et al. [71] reported that pore pressure in shale reservoirs is typically greater than 20 MPa (see their Tables 5 and 6). In another study, Gherabati et al. [72] indicated that for Eagle Ford shale the pore pressure varies from near 14 to about 97 MPa.

*6.2. Critical gas saturation*

One of the principle goals of enhanced gas recovery is reducing critical gas saturation $S_{gc}$ in previously swept zones of a gas reservoir [73]. In this study, following Katz and Thompson [74], we estimated the value of $S_{gc}$ from the inflection point of MICP curve. Although we obtained reasonably well $k_{rg}$ estimations for half of samples studied here, one may estimate $S_{gc}$ from depth of penetration plot [75], threshold gas displacement [76], or MICP in combination with electrical resistivity measurements [77]. It is also well documented in the literature that the value of $S_{gc}$ depends on several factors, such as pore connectivity, capillary number and flow rate, sample size, and wetting characteristics [21,32,73].

Generally speaking, the higher the pore connectivity, the smaller the volume of trapped gas and, consequently, the smaller the critical gas saturation. A rough approximation from bond percolation theory is $S_{gc} \approx 1.5/Z$ [32] in which $Z$ is the average pore coordination number. However, results of Iglauer et al. [78] and Ghanbarian et al. [49] demonstrated that $1.5/Z$ typically overestimates $S_{gc}$ in porous media. One should expect large $S_{gc}$ values in media with significant fraction of dead-end pores or isolated pore clusters. Using X-ray tomography, Iglauer et al. [78] showed that the size distributions for the non-wetting phase in two sandstones conformed to power-law cluster-size distribution from percolation theory with exponents (e.g., 2.053) very close to the theoretical universal value of 2.18 [32].

Capillary number and flow rate also influence critical saturation for relative permeability. Early experiments by Larson et al. [79] and Chatzis and Morrow [80] revealed that critical saturation for relative permeability is approximately constant at low capillary numbers. However, at some capillary number a knee occurs in the critical saturation-capillary number curve and critical saturation starts to decrease (see Fig. 3.21 in Lake et al. [73]).

The critical gas saturation is also scale-dependent. Within the percolation theory framework, one should expect $S_{gc}$ to increase as system size increases. The finite-size scaling theory of Fisher [81] shows that the finite size of a medium causes a shift in the percolation threshold and/or critical gas saturation.

The effect of contact angle and wettability of porous rocks has been known for decades (see e.g., Anderson [82]). In contrast to homogenously-wet rocks such as sandstones, shales are typically heterogeneously-wet meaning that the contact angle in organic patches might be significantly different from that in the inorganic matrix. Huang et al. [16] found that residual gas saturation in their mixed-wet networks was less than that in completely wat-wet media. They also reported that gas relative permeability was higher in their mixed-wet systems because of the presence of hydrocarbon-wet pore bodies and throats. Huang et al. [16] stated that, " … the trapping of gas observed in the imbibition simulation is minimized in the organic region, because in the drainage process, the gas in the large pores is displaced first. This can also explain the lower residual gas saturation in the mixed-wet system."

Further investigations are required to better understand mechanisms causing gas trapping in nano-scale pores and to more accurately estimate critical gas saturation for shales and tight porous rocks under two-phase flow conditions.

**7. Conclusions**

In this study, we applied concepts from the effective-medium approximation and universal power-law scaling from percolation theory to develop a theoretical gas relative permeability $k_{rg}$ model in shales and tight porous rocks. For this purpose, we assumed that gas transport is mainly controlled by two mechanisms contributing in parallel: (1) hydraulic flow and (2) molecular flow. Our proposed model estimates $k_{rg}$ from pore space properties such as pore size distribution, pore connectivity (reflected in critical gas saturation) and porosity as well as gas characteristics e.g., gas type, viscosity, temperature and molar mass. To evaluate the proposed theoretic approach, we compared model estimations with six experimental measurements and two pore-network simulations from the literature. Results showed that $k_{rg}$ was satisfactorily estimated for the entire range of gas saturation in shales and tight porous rocks except Case VI which might have some microfractures.

**Acknowledgment**

The authors acknowledge Rasoul Nazari Moghaddam, Delft University of Technology, and Hassan Dehghanpour, University of Alberta, for sharing the experimental data used in this study, as well as Wenhui Song, China University of Petroleum (East China), and Xiukun

Wang, China University of Petroleum at Beijing, for their fruitful personal communications. BG is grateful to Kansas State University for supports through faculty startup funds.

[46] Ghanbarian-Alavijeh, B., Hunt AG. Comparison of the predictions of universal scaling of the saturation dependence of the air permeability with experiment. Water Resour Res 2012;48:W08513.

[47] Ghanbarian B, Hunt AG, Skinner TE, Ewing RP. Saturation Dependence of Transport in Porous Media Predicted By Percolation and Effective Medium Theories. Fractals 2015;23:1540004. doi:10.1142/S0218348X15400046.

[48] Ghanbarian B, Daigle H, Hunt AG, Ewing RP, Sahimi M. Gas and solute diffusion in partially saturated porous media: Percolation theory and Effective Medium Approximation compared with lattice Boltzmann simulations. J Geophys Res Solid Earth 2015;120. doi:10.1002/2014JB011645.

[49] Ghanbarian B, Hamamoto S, Kawamoto K, Sakaki T, Moldrup P, Nishimura T, et al. Saturation-dependent gas transport in sand packs: Experiments and theoretical applications. Adv Water Resour 2018;122:139–47. doi:10.1016/j.advwatres.2018.10.011.

[50] Daigle H, Ezidiegwu S, Turner R. Determining Relative Permeability In Shales By Including The Effects Of Pore Structure On Unsaturated Diffusion And Advection. SPE Annu Tech Conf Exhib 2015. doi:10.2118/175019-MS.

[51] Ojha S. P., Misra, S., Tinni, A., Sondergeld, C., Rai C. Relative permeability estimates for Wolfcamp and Eagle Ford shale samples from oil, gas and condensate windows using adsorption- desorption measurements. Fuel 2017;208:52–64.

[52] Li J, Chen Z, Wu K, Li R, Xu J, Liu Q, et al. Effect of water saturation on gas slippage in tight rocks. Fuel 2018;225:519–32. doi:10.1016/j.fuel.2018.03.186.

[53] Bennion DB, Bachu S. Permeability and relative permeability measurements at

**Table 1**
Salient properties of the eight cases used in this study.

| Case | Reference | Method | Remark | $\phi$ | $S_{gr}$ | $S_{gc}$ |
|---|---|---|---|---|---|---|
| I | Bennion and Bachu [53] | Exp | Calmar shale | 0.039 | 0 | $0^*$ |
| II | Bennion and Bachu [53] | Exp | Colorado shale | 0.044 | 0 | 0.30 |
| III | Dacy [54] | Exp | Tight-gas sand | 0.175 | 0 | 0.08 |
| IV | Yassin et al. [55] | Exp | Tight-gas siltstone | 0.062 | 0 | 0.26 |
| V | Nazari Moghaddam and Jamiolahmady [56] | Exp | Eagle Ford Shale | 0.135 | 0 | 0.13 |
| VI | Nazari Moghaddam and Jamiolahmady [56] | Exp | Pierre Shale | 0.318 | 0 | $0.08^*$ |
| VII | Song et al. [17] | PNM | Shale | 0.150 | 0 | 0.05 |
| VIII | Wang and Sheng [58] | PNM | Shale | 0.075 | 0 | $0.43^*$ |

Exp: experiment, PNM: pore-network model, $\phi$: porosity (fractional), $S_{gr}$: residual gas saturation, $S_{gc}$: critical gas saturation
* $S_{gc}$ was determined from the measured/simulated $k_{rg}$ curve.

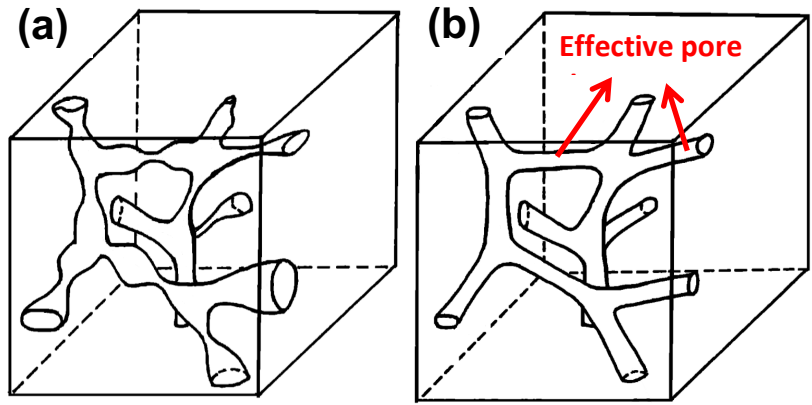

**Fig. 1.** Three-dimensional schematic of a heterogeneous porous rock (a) replaced by a homogeneous pore network (b) with an effective pore size (a modified version of Doyen [24]).

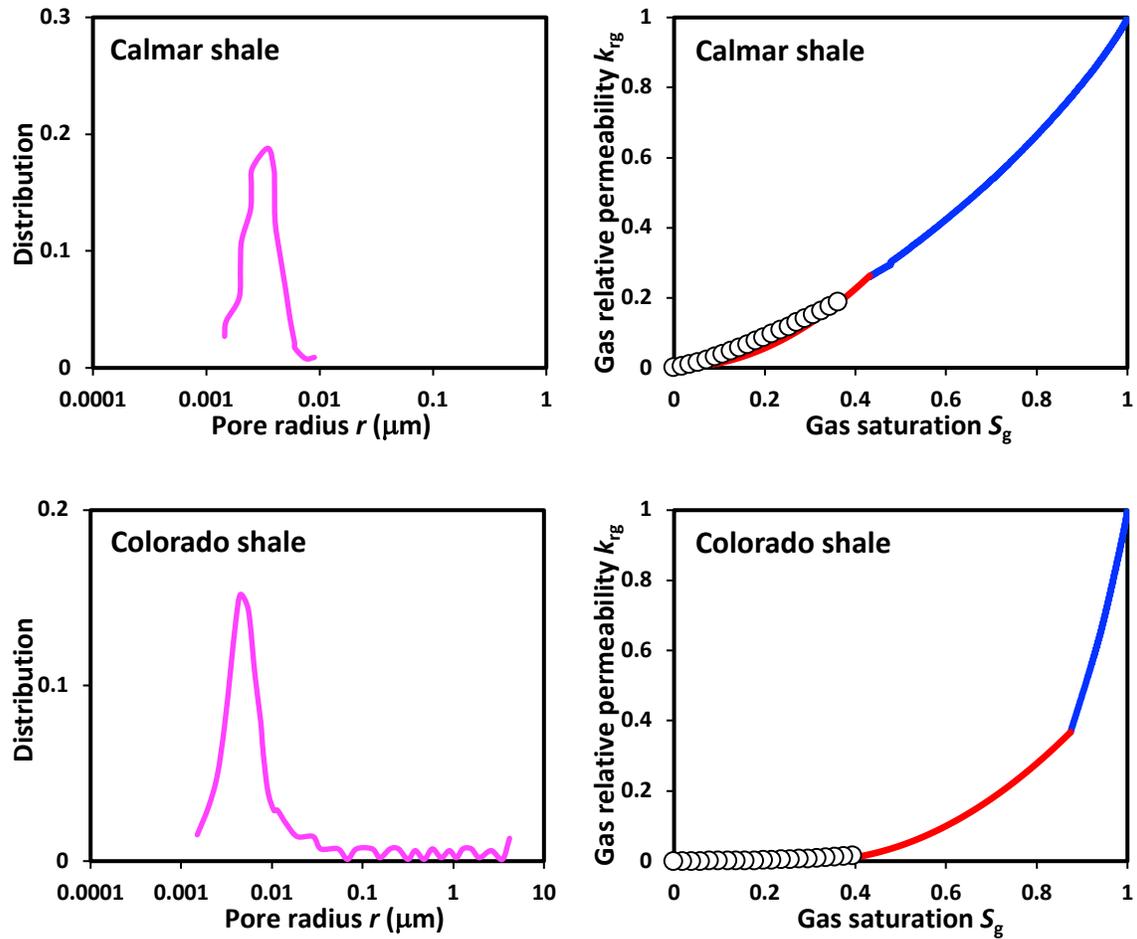

**Fig. 2.** The pore size distribution, derived from the measured mercury intrusion porosimetry curves, (on the left) and the estimated gas relative permeability curve (on the right) for Calmar and Colorado shales from Bennion and Bachu [53]. The blue and red lines represent the results of effective-medium approximation (Eq. 11) and universal power-law scaling from percolation theory (Eq. 12), respectively. Unfilled circles represent $k_{rg}$ measured at $p$ = 12.25 MPa (Calmar) and 20 MPa (Colorado). Note that $S_{gc} = 0$ for Calmar was determined from the actual $k_{rg}$ measurements, while for Colorado $S_{gc} = 0.30$ was estimated from the inflection point of the measured MICP curve.

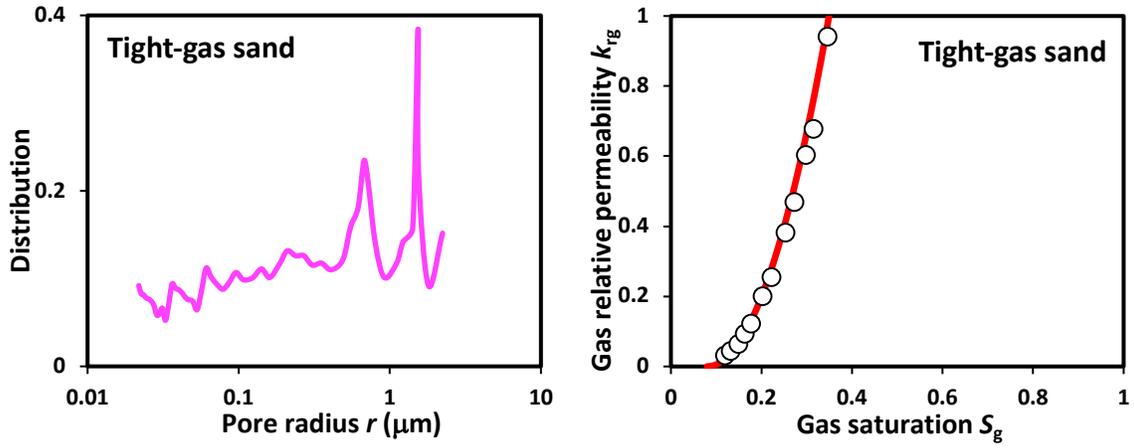

**Fig. 3.** The pore size distribution, derived from the air-brine capillary pressure curve measured via centrifuge, (on the left) and the estimated gas relative permeability curve (on the right) for tight-gas sand from Dacy [54]. For this dataset, the proposed model reduced to the power-law scaling from percolation theory (Eq. 12) represented by the red line. Unfilled circles represent $k_{rg}$ measured at $p = 6.9$ MPa. Note that $S_{gc} = 0.08$ was estimated from the inflection point of the measured MICP curve.

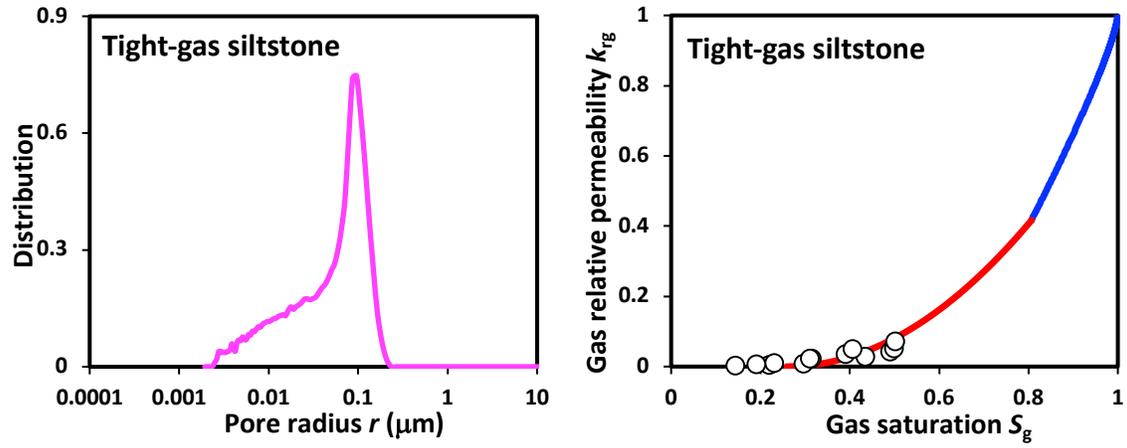

**Fig. 4.** The pore size distribution, derived from the measured mercury intrusion porosimetry curve, (on the left) and the estimated gas relative permeability curve (on the right) for tight-gas siltstone from Yassin et al. [55]. The blue and red lines represent the results of effective-medium approximation (Eq. 11) and universal power-law scaling from percolation theory (Eq. 12), respectively. Unfilled circles denote $k_{rg}$ measured at $p = 6.9$ MPa. Note that $S_{gc} = 0.26$ was estimated from the inflection point of the measured MICP curve.

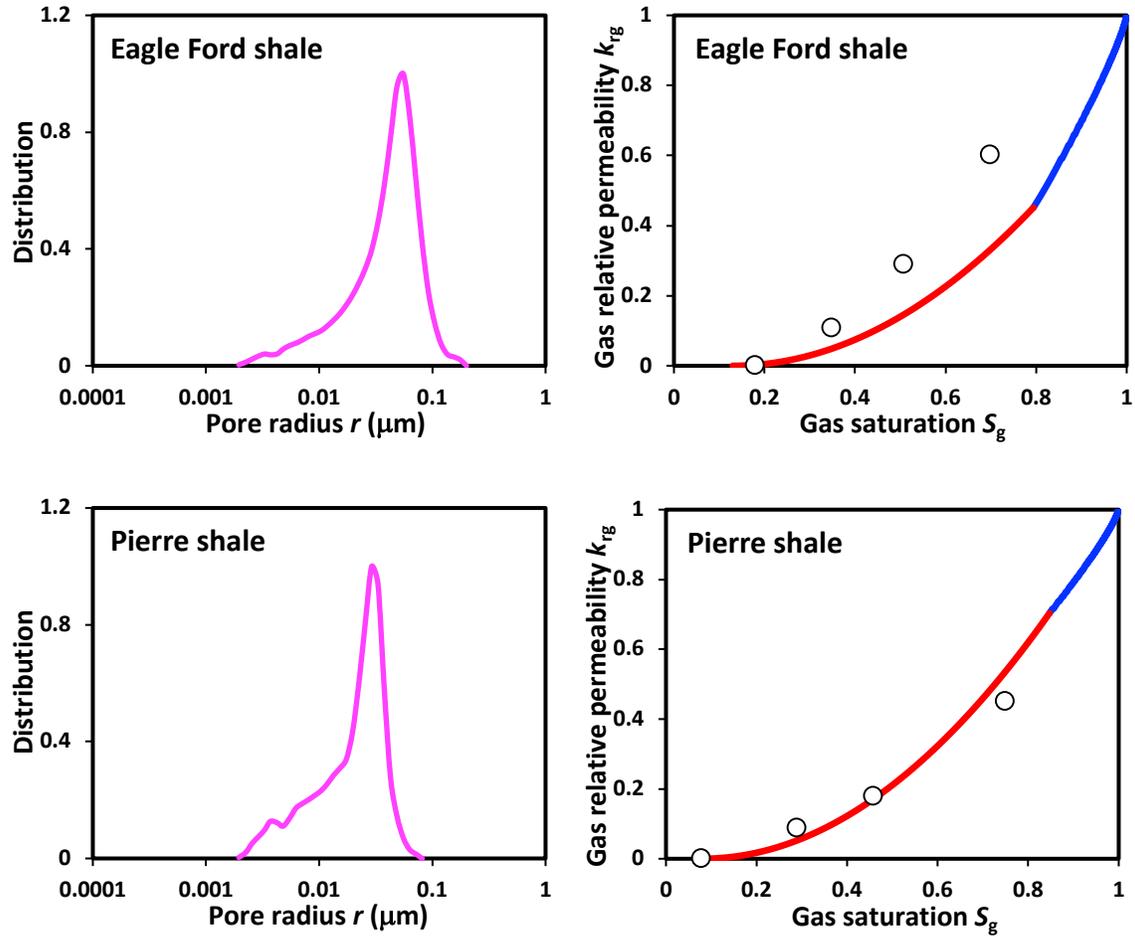

**Fig. 5.** The pore size distribution, derived from the measured mercury intrusion porosimetry curve, (on the left) and the estimated gas relative permeability curve (on the right) for Eagle Ford and Pierre shale samples from Nazari Moghaddam and Jamiolahmady [56]. The blue and red lines represent the results of effective-medium approximation (Eq. 11) and universal power-law scaling from percolation theory (Eq. 12), respectively. Unfilled circles denote $k_{rg}$ measured at $p = 10.3$ MPa. Note that $S_{gc} = 0.13$ for Eagle Ford was estimated from the inflection point of the measured MICP curve, while for Pierre $S_{gc} = 0.08$ was determined from the actual $k_{rg}$ measurements.

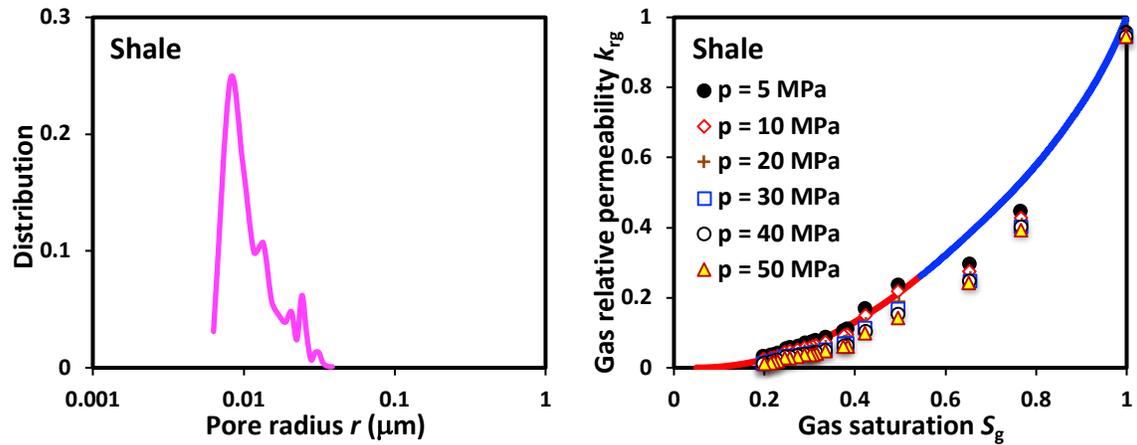

**Fig. 6.** The pore size distribution, constructed from two-dimensional SEM images of the organic-rich shale sample, (on the left) and the estimated gas relative permeability curve at $p = 25$ MPa (on the right) from Song et al. [17]. The blue and red lines represent the results of effective-medium approximation (Eq. 11) and universal power-law scaling from percolation theory (Eq. 12), respectively. Symbols denote $k_{rg}$ simulated via the pore-network model at various $p$ values. Note that $S_{gc} = 0.05$ was determined from the fraction of dead-end pores in the network.

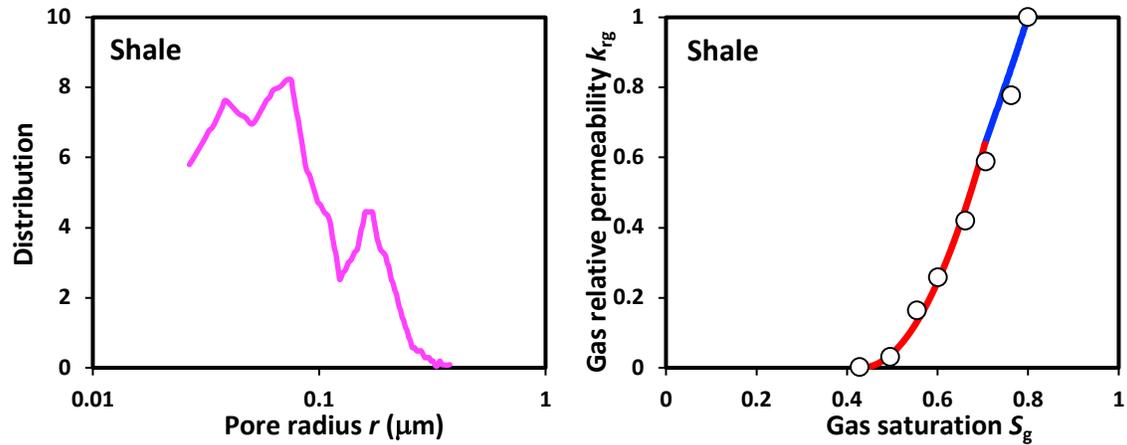

**Fig. 7.** The pore size distribution, synthetically generated from a Berea sandstone to mimic the actual pore size distribution in a shale sample, (on the left) and the estimated gas relative permeability curve at $p$ = 6.9 MPa (on the right) for the pore-network simulations from Wang and Sheng [58]. The blue and red lines represent the results of effective-medium approximation (Eq. 11) and universal power-law scaling from percolation theory (Eq. 12), respectively. Note that $S_{gc}$ = 0.43 was determined from the $k_{rg}$ simulations.